\documentclass[1p,review]{elsarticle}
\pdfoutput=1 % if your are submitting a pdflatex (i.e. if you have
\usepackage[english]{babel}
\usepackage{xcolor}
\usepackage{subcaption}
\usepackage{wrapfig}
\usepackage{multirow}
\usepackage{amsmath}
\usepackage{siunitx}
\usepackage{natbib}
\usepackage{url}

\usepackage{breakurl}
\usepackage[breaklinks]{hyperref}
\title{\boldmath Implementation and performances of the IPbus protocol\\
for the JUNO Large-PMT readout electronics}
\author[a]{Riccardo~Triozzi}
\author[a,b]{Andrea~Serafini\corref{cor1}}
\cortext[cor1]{Corresponding author: andrea.serafini@infn.it}
\author[b]{Marco~Bellato}
\author[b]{Antonio~Bergnoli}
\author[a,b]{Matteo~Bolognesi}
\author[a,b]{Riccardo~Brugnera}
\author[a]{Vanessa~Cerrone}
\author[c]{Chao~Chen}
\author[d]{Barbara~Clerbaux}
\author[a]{Alberto~Coppi}
\author[b]{Daniele~Corti}
\author[b]{Flavio~dal~Corso}
% ...
\cortext[cor2]{Now at INFN Laboratori Nazionali del Sud, Italy}
\author[a,b]{Chiara~Sirignano}
\author[l]{Monica~Sisti}
\author[b]{Luca~Stanco}
\author[i,j]{Virginia~Strati}
\author[h,g]{Francesco~Tortorici}
\author[h,g]{Cristina~Tuv\'e}
\author[m]{Carlo~Venettacci}
\author[g]{Giuseppe~Verde}
\author[p]{Lucia~Votano}
\address[g]{INFN Sezione di Catania, Catania, Italy}
\address[h]{Universit\`a di Catania, Dipartimento di Fisica e Astronomia, Catania, Italy}
\address[i]{INFN Sezione di Ferrara, Ferrara, Italy}
\address[j]{Universit\`a degli Studi di Ferrara, Dipartimento di Fisica e Scienze della Terra, Italy}
\address[k]{INFN Sezione di Milano e Universit\`a di Milano, Dipartimento di Fisica, Milano, Italy}
\address[l]{INFN Sezione di Milano Bicocca, e Universit\`a di Milano Bicocca, Dipartimento di Fisica, Milano, Italy}
\address[m]{INFN Sezione di Roma Tre e Universit\`a di Roma Tre, Dipartimento di Matematica e Fisica, Roma, Italy}
\address[n]{INFN, Sezione di Milano Bicocca e Politecnico di Milano, Dipartimento di Energetica, Milano, Italy}
\address[o]{INFN Sezione di Perugia e Universit\`a di Perugia, Dipartimento di Chimica, Biologia e Biotecnologie, Perugia, Italy}
\address[p]{Laboratori Nazionali dell'INFN di Frascati, Italy}

\begin{document}

% ---------------------------------------------------
% ABSTRACT, KEYWORDS
% ---------------------------------------------------
\begin{abstract}
The Jiangmen Underground Neutrino Observatory (JUNO) is a large neutrino detector currently under construction in China. Thanks to the tight requirements on its optical and radio-purity properties, it will be able to perform leading measurements detecting terrestrial and astrophysical neutrinos in a wide energy range from tens of keV to hundreds of MeV. A key requirement for the success of the experiment is an unprecedented  3\% energy resolution, guaranteed by its large active mass (20~kton) and the use of more than 20,000 20-inch photo-multiplier tubes (PMTs) acquired by high-speed, high-resolution sampling electronics located very close to the PMTs.
As the Front-End and Read-Out electronics is expected to continuously run underwater for 30~years, a reliable readout acquisition system capable of handling the timestamped data stream coming from the Large-PMTs and permitting to simultaneously monitor and operate remotely the inaccessible electronics had to be developed.
In this contribution, the firmware and hardware implementation of the IPbus based readout protocol will be presented, together with the performances measured on final modules during the mass production of the electronics.
\end{abstract}


\maketitle
\newpageafter{title}

% ---------------------------------------------------
% INTRODUCTION
% ---------------------------------------------------
\section{Introduction}
\label{sec:intro}
The Jiangmen Underground Neutrino Observatory (JUNO) is a 20 kton multi purpose underground liquid scintillator detector currently under construction in the Guangdong Province in South China \cite{bib:juno:phys-det}. The main goal of the JUNO experiment is the determination of the Neutrino Mass Ordering (NMO), which will be resolved by analyzing the oscillation pattern of the electron anti-neutrinos produced in the nuclear fissions of the 52.5 km-distant Yangjian and Taishan Nuclear Power Plants. Within six year of data-taking, JUNO is expected to determine the NMO at a $3\sigma$ significance \cite{bib:juno:phys-det}, simultaneously measuring the $\Delta m^2_{31}$, $\Delta m^2_{21}$, $\sin^2 \theta_{12}$ oscillation parameters to a sub percent World leading precision \cite{bib:JUNO:2022mxj}. Besides its main ambitious goal, JUNO’s extensive physics program includes studies of neutrinos from the Sun, the atmosphere, supernovae, and planet Earth, as well as explorations of physics beyond the Standard Model \cite{bib:juno:phys-det}.\\
\indent In order to fulfill its ambitious goals, JUNO aims to achieve a better than 1\% energy linearity and a 3\% energy resolution at 1 MeV \cite{bib:juno:calib_strategy}. This unprecedented resolution will be reached by featuring a dual calorimetry system \cite{bib:dual_cal} consisting of 17,612 20-inch Large-PMTs \cite{bib:juno:lpmt_test} and a secondary 25,600 3-inch Small-PMTs system \cite{bib:juno:spmt_test} in the central detector, resulting in a remarkable 78\% photon sensor coverage \cite{bib:juno:phys-det}. Additionally, 2,400 20-inch Large-PMTs will be installed in the 35-kton water pool surrounding the detector, acting as part of the muon veto system.
While for Small-PMTs only the temporal and deposited charge information will be gathered for each signal, for Large-PMTs the entire waveforms will be collected and stored, resulting in a challenging amount of data to manage. In JUNO, the design trigger rate during data-taking is 1~kHz \cite{bib:juno:phys-det}; this value has been used as target for the development of the readout electronics. A reliable readout acquisition system capable of handling the synchronized data stream coming from the Large-PMTs is a strict requirement for the success of the experiment. 

The signal acquired by the 20,012 Large-PMTs used in the central detector and the water tank will be digitized by 7000 Global Control Unit (GCU) boards. With the aim of ensuring minimum electrical noise, the GCUs will be installed underwater in the vicinity of the PMTs. After the filling of the experiment and during data taking, the GCU boards will be not accessible anymore. It is therefore crucial to design a protocol permitting to monitor and operate remotely the electronic boards for the entire duration of the experiment. These requirements are addressed through the development of a custom designed readout electronics \cite{bib:el_paper, bib:juno:elec_ReD} and the implementation of an IPbus-based protocol \cite{bib:ipbus} for the simultaneous transmission of data and the slow control of the GCU boards' parameters through Ethernet. In order to test the performances of the IPbus implementation in the JUNO Data AcQuisition (DAQ) streams, extensive rate and bandwidth measurements were collected using two different facilities. Results are reported in the following sections.

% ---------------------------------------------------
% SETUP OVERVIEW
% ---------------------------------------------------
\section{Setup overview}
\label{sec:setup_overview}
\subsection{JUNO large PMT electronics}
\label{sec:design}
\begin{figure}[htbp]
\centering
  \includegraphics[width=1.\columnwidth]{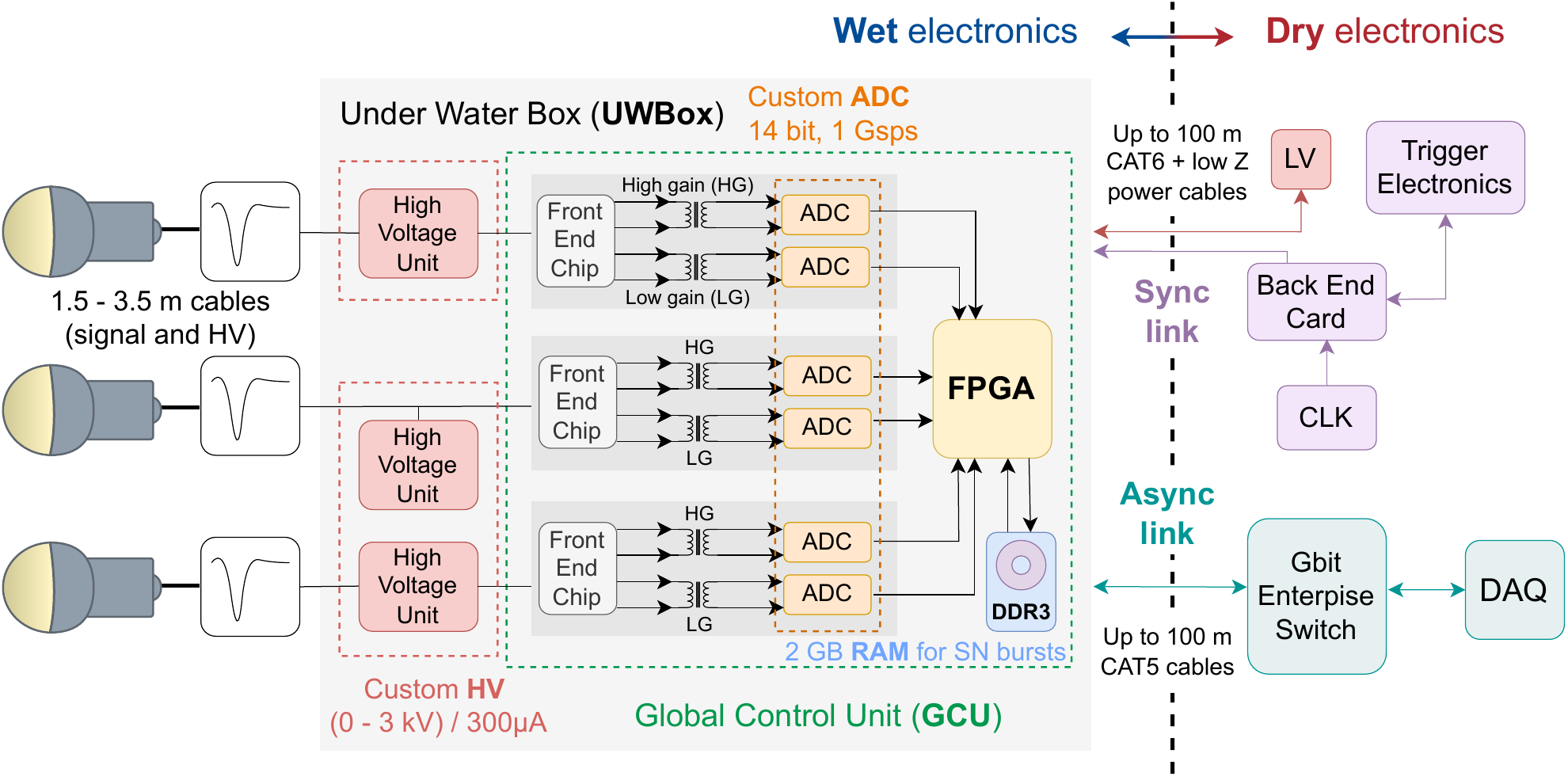}
  \caption{\label{fig:1f3:scheme}JUNO large PMT electronics Read-Out
           electronics scheme. A description of the different parts is
           given in the text.}
\end{figure}
A scheme of the JUNO large-PMT electronics is given in Figure~\ref{fig:1f3:scheme}~\cite{bib:el_paper}.
The design is an optimization of previous developments~\cite{bib:juno:elec_ReD}. 
The full electronics chain is composed of two parts: the \emph{front-end} (FE), or \emph{wet}, electronics~\cite{bib:Marini:phd} located very close to the LPMT output, inside the JUNO Water Pool; and the \emph{dry} electronics, installed in the electronics rooms of the JUNO underground laboratories, which consists of the \emph{back-end} (BE), or trigger, electronics and the data acquisition (DAQ) system. 
The FE electronics will be installed on the JUNO Steel Truss structure, inside a stainless steel, water-tight box, the so-called Under Water Box (UWbox).
Three PMT output signals are fed to one UWbox which contains:
\begin{itemize}
\item three High Voltage Units (HVU): programmable modules which provide the bias voltage to the PMT voltage divider. Each HVU independently powers one large PMT. The HVUs are mounded on a custom PCB, the splitter board, that provides mechanical stability to the modules, and decouples the PMT signal current from the high voltage.
\item a Global Control Unit (GCU): a motherboard incorporating the Front-End and Readout electronics components. The three PMT signals reaching the GCU are processed through independent readout chains.
\end{itemize}
%
% The PMT analog signal reaching the GCU, is duplicated and processed,
% in parallel, by a low-gain and high-gain transimpedance amplifiers.
The PMT analog signal reaching the GCU is processed by a custom Front-End Chip (FEC) which provides two outputs with a low and high gain.
The signal is further converted to a
digital waveform by a 14 bit, 1 GS/s, custom FADC. 
The usage of two FADCs per readout channel has been driven by the design
requirement of providing a wide dynamic range in terms of
reconstructed photo-electrons (pe): from 1 pe to 100 pe with a 0.1 pe
resolution (high gain FADC) and from 100 pe to 1000 pe (low gain FADC) with
a 1 pe resolution~\cite{bib:juno:yb}.

A Xilinx Kintex-7 FPGA (XC7K325T) is the core of the GCU and allows to further process the digital signal and temporarily store it in a
local memory buffer before sending it to the data acquisition (DAQ). The Kintex-7 processing unit is responsible for the onboard generation of a trigger request (referred to as local trigger) whenever the digitized signal overpass the preset threshold, the charge reconstruction and the timestamp tagging of the acquired waveforms.
Besides the local memory available in the readout-board FPGA, a 2~GBytes
DDR3 memory is available and used to provide a larger memory buffer in the
exceptional case of a sudden increase of the input rate, which overruns
the current data transfer bandwidth between the FE electronics and the DAQ.
An additional \mbox{Spartan-6}
FPGA (XC6SLX16) is available on the same motherboard.
It implements a 2-port Ethernet hub and a RGMII interface between the
PHY network chip and the Spartan-6 and it also interconnects the Spartan-6
and the Kintex-7. The Spartan-6 FPGA provides an important
failsafe reconfiguration feature of the Kintex-7 by means of a virtual
JTAG connection over the IPbus, removing the need of a dedicated JTAG
connector and cable.

The BE electronics is composed of the following active elements:
\begin{itemize}
\item the Back End Card (BEC) with the Trigger and Time Interface Mezzanine
      (TTIM)
\item the Reorganize and Multiplex Units (RMU) and the Central Trigger
      Unit (CTU), which are part of the Trigger Electronics
      (see Figure~\ref{fig:1f3:scheme}).
\end{itemize}

The PMTs are connected to the UWbox electronics with a $50~\Omega$,
coaxial cable. 
The electronics inside the UWbox has two independent connections to
the BE electronics:
a so-called ’synchronous link’ (S-link), which provides the clock
and synchronization to the boards and handles the trigger primitives,
and an ’asynchronous link’ (A-link) which is fully dedicated to the DAQ
and slow-control. These connections are realized using commercially available
CAT-5 and CAT-6 Ethernet cables for the A-link and S-link, respectively.
The length of the cables is fixed to about 1.5~m between the PMTs and
the UWbox and will have a variable length between 30~m and 100~m
for the connection between the UWbox and the BE electronics in the electronics
rooms.
An additional, low-resistance, power cable will be used to bring power
to the electronics inside the UWbox.

The large PMT electronics can run with a centralized ’global trigger’ mode,
where the information from the single ‘fired’ PMTs is collected and processed
in the Central Trigger Unit (CTU). The latter validates the trigger based on
a simple PMT multiplicity condition or a more refined topological distribution
of the fired PMTs in JUNO. Upon a trigger request, validated waveforms are
sent to the DAQ event builder through the A-link. The IPBus Core protocol \cite{bib:ipbus} is
used for data transfer, slow control monitoring, and electronics configurations.

An alternative scheme is possible where all readout boards send their locally
triggered waveforms to the DAQ, independently of each other. With this
approach, all the digitized waveforms, including those generated by dark
noise photoelectrons, will be sent to the DAQ.

\subsection{The JUNO test facilities}
\label{sec:facilities}
% [Commenti: la metterei in questo modo... Si sono scelte due diverse facilities. A Legnaro si è fatto un piccolo mock up dell'esperimento in modo da poter testare l'intera catena di presa dati dal PMT fino al trasferimento della waveform su PC. Questo mock up ha 48 PMT e via dicendo [cita paper vanessa-kate]. Per testare le performances dell'elettronica di lettura su larga scala si è invece deciso di usare la facility di Kunshan. Qui è dove avviene l'assemblaggio delle GCU. Il setup prevede 300 e sbrisga GCU connesse in parallelo ad un server simile a quello che verrà impiegato in JUNO che consiste di 24 cores e 48 threads [cita paper bea-alberto]. Questo ha permesso di valutare estensivamente e in modo realistico ecc. ecc. Toglierei la figura da qui e me la giocherei nella sezione sull'implementazione dell'IPbus]

To test the large PMTs Read-Out electronics and the IPbus performances in a realistic scenario, two different setups were assembled. A small JUNO mock-up system exploiting the full electronics chain from the PMTs to the DAQ server was set up at the Legnaro National Laboratories (LNL) of the Italian National Institute for Nuclear Physics (INFN) \cite{bib:LNL, bib:Cerrone:Bachelor, bib:Triozzi:Bachelor} in Legnaro (Padua, Italy). This allowed to test the correct interplay among the different read-out components and the flawless implementation of the IPbus protocol. A second apparatus was set up in a dedicated test facility in Kunshan (China), where the mass production and assembling of the GCUs is currently underway \cite{bib:juno:kunshan_tests, bib:Coppi:Bachelor}. The large-scale setup built in Kunshan allowed to evaluate extensively and realistically the electronics performances on a scale comparable to that envisaged for the JUNO experiment.

\paragraph{LNL setup}
The medium size setup installed at LNL features 48 PMTs, associated with as many independent channels. The 2-inch PMTs\footnote{Philips XP2020 photomultiplier tubes} are submerged in a cylindrical container holding 17 liters of liquid scintillator. A total of 16 GCUs are connected to a single BEC conveying the synchronous information to an RMU coupled to a CTU. The DAQ system is installed in a dedicated server running Ubuntu 18, featuring 32 GB of RAM and 2x Intel Xeon Silver 2.2 GHz CPUs, for a total of 20 cores and 40 threads. 
An extensive and technical description of the apparatus can be found in \cite{bib:Cerrone:Bachelor, bib:juno:lnl_tests}.
% In particular, the electronics chain scheme is shown in Figure \ref{fig:lnl-chain}. 

% \begin{figure}[hbtp] \centering
%   \includegraphics[width=0.9\columnwidth]{figures/catenaLNL_final.pdf}
%   \caption{\label{fig:lnl-chain}Scheme of the LNL electronics chain.}
% \end{figure}

% The measurements considered for this paper were collected using 11 GCUs at most, resulting in 33 PMTs and thus acquisition channels that collect signals asynchronously. The signal flow is the following:
% \begin{enumerate}
%     \item the Analog-to-Digital Unit (ADU) in the GCU digitizes and doubles the analog signal produces for each channel: one signal copy is registered with the corresponding GCU timestamp into a L1 cache and the other copy is analyzed with the trigger algorithm;
%     \item if the signal is consistent with a certain threshold, a trigger request is sent to the BEC (via synchronous link), which can validate the trigger decision sending back the same timestamp to all involved GCUs. The considered measurements were collected using the \textit{external trigger mode}: the trigger request is considered valid when it coincides within a predefined time window with an external trigger generated by the HP Agilent Keysight 8116A pulser and it is sent directly to the BEC;
%     \item finally, the signals with the selected timestamp are transferred (via asynchronous link) from the L1 cache to the First-In-First-Out (FIFO) unit and then moved to the PC through an Ethernet switch, where the DAQ stores the data.
% \end{enumerate}
%
\paragraph{Kunshan setup}
\label{subsec:exp_setup_Kunshan}
The Kunshan test facility hosts a large-scale setup where a total of 344 GCUs can be managed simultaneously. Groups of 40 GCUs are handled by a total of 9 BECs, connected to as many RMUs controlled by a single CTU. The DAQ software is installed on a Intel Xeon Gold based server running on CentOS7, featured with 192~GB of RAM and, 2x 2.7GHz processors, for a total of 24 cores and 48 threads. A detailed description of the Kunshan test facility can be found in \cite{bib:juno:kunshan_tests}. 

% ---------------------------------------------------
% IPBUS IMPLEMENTATION
% ---------------------------------------------------
\section{IPbus implementation in the data acquisition stream}
\label{sec:ipbus}
One of the main challenges for the JUNO Large-PMTs Read-Out chain is the possibility to acquire and transfer data in parallel to the remote monitoring and control of the electronics. After the filling of the JUNO detector, the FE will be inaccessible for the entire duration of the data-taking (expected to run for 30 years \cite{bib:juno:phys-det}). It is therefore imperative to implement a protocol assuring a reliable transmission of data, while allowing the asynchronous request and remote modification of acquisition parameters. In this respect, the IPbus suite of software and firmware implements a reliable high-performance control link specifically suited for particle physics electronics \cite{bib:ipbus}, as demonstrated by its successful employment, among others, in the CMS \cite{ipbus:cms}, ATLAS \cite{ipbus:atlas} and ALICE \cite{ipbus:alice} experiments. 

IPbus is a hardware and firmware solution that communicates over Ethernet using UDP/IP. It consists of (i) a firmware module implementing the IPbus protocol within end-user hardware (e.g. JUNO's Kintex-7 and Spartan-6 FPGAs), (ii) a micro Hardware Access Library (uHAL) providing an end-user C++/Python library for read/write operations on IPbus and (iii) a software application called ControlHub, which mediates simultaneous hardware access from multiple uHAL clients. While the UDP protocol does not include any native reliability mechanism, the use of ControlHub assures the duplication and re-ordering of any lost IPbus UDP packet, providing a reliability mechanism at software level.
\subsection{Hardware and firmware implementation}
The IPBus protocol is deployed to the GCU via an FPGA core provided in the IPBus suite. This core provides the following modules: (i) a core protocol decoder and (ii) a bus master that controls the interface to several IPBus slaves (e.g. onboard trigger manager, L1-cache, IPbus DAQ, UART-based high voltage unit, I$^2$C-based temeperature sensors, DDR3 memory). Each slave contains a set of registers that can be read and written by IPBus transactions. The addresses, sub-addresses and masks of these slaves registers can be specified by Extensible Markup Language (XML) files.

\begin{figure}[hbtp] \centering
  \includegraphics[width=1.\columnwidth]{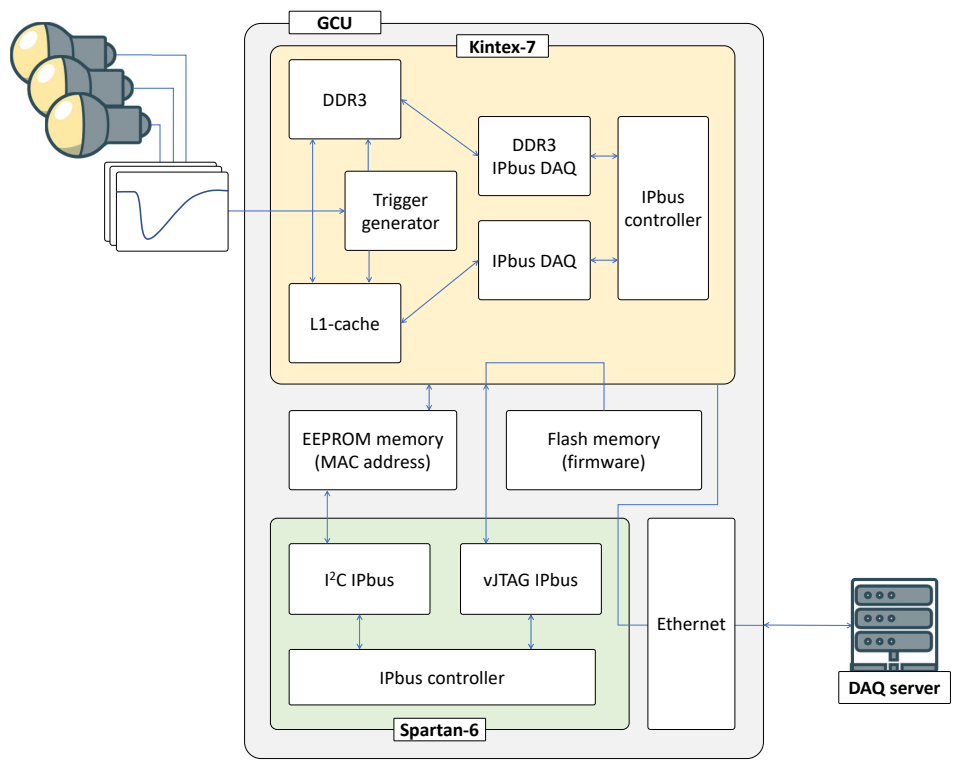}
%   \caption{\label{fig:lnl-chain}Scheme of the LNL electronics chain.}
  \caption{\label{fig:gcu-scheme}Scheme of the IPbus implementation within the GCU. The tasks and cores of the Kintex-7 (in yellow) and Spartan-6 (in green) FPGAs are indicated in the white boxes.}
\end{figure}

In the Kintex7 FPGA, the IPbus master-slave interface is mainly used for data acquisition (see Figure \ref{fig:gcu-scheme}). Once that an analog waveform coming from a PMT has been digitized by the ADC, the digital data follows a three-fold path. The data is sent to (i) a circular buffer called L1-cache capable of storing 32 $\mu$s of raw data per channel, (ii) a 2 GB DDR3 memory accommodating high trigger rate scenarios and (iii) a trigger algorithm which extracts the timing information and sends the obtained timestamp to the BE, the DDR3 packager and the L1-cache.
The DDR3 memory is designed to run in self triggering mode (no validation from the CTU is required). After receiving the correct timestamp, a packager module encapsulates the digitized event within an header and a trailer and sends it to a DDR3 controller module which addresses the memory as a circular buffer. Whenever the DDR3 content is requested by the DAQ via an IPbus transaction, the writing operations are blocked and the DDR3 is emptied out by the IPbus DDR3 DAQ module. Once the whole content is read, the writing restarts automatically.
The L1-cache is instead designed to work in external triggering mode (trigger requests needs to be validated by the CTU), but can be programmed to work in self triggering mode via IPbus if necessary. As soon as a timestamped trigger arrives to the L1-cache, a fixed data payload is extracted from the buffer, packetized within an header and trailer, and sent to the IPbus DAQ module. The IPBus DAQ module presents a “funnel” asynchronous First In, First Out (FIFO) buffer of 2$^{13}$ bytes able to contain four waveform packets, which can be read via IPbus transactions. 
The IPbus controller core, that provides the interfaces for the IPbus DAQ and IPbus DDR3 DAQ slaves, encapsulates the payload data into UDP network compliant packets and sends the data to the server via the Ethernet MAC module.

In the Spartan-6 FPGA, the IPbus master-slave interface is instead mainly used for the MAC address assignment and for permitting the remote reprogramming of the Kintex-7's firmware  (Figure \ref{fig:gcu-scheme}). An I$^2$C IPbus slave is used for writing/reading a small EEPROM memory storing the MAC address. A dedicated on-board bus then transfers the MAC address information from the Spartan-6 to the Kintex-7 FPGAs.
A virtual JTAG (vJTAG) IPBus slave is instead used to expose the Kintex-7 JTAG hardware access to the network (see Section \ref{jtag}). The vJTAG IPbus slave receives from the server the JTAG commands encapsulated in a standard network transport layer, extracts the JTAG instructions and then uses a dedicated bus to exchange the JTAG commands to/from the Kintex-7.

% \begin{figure}[hbtp] \centering
%   \includegraphics[width=0.9\columnwidth]{figures/scheme_IPbus.pdf}
%   \caption{\label{fig:ipbus-scheme}Scheme of IPbus??????.}
% \end{figure}

\subsection{Software implementation}
\begin{figure}[hbtp] \centering
  \includegraphics[width=1.\columnwidth]{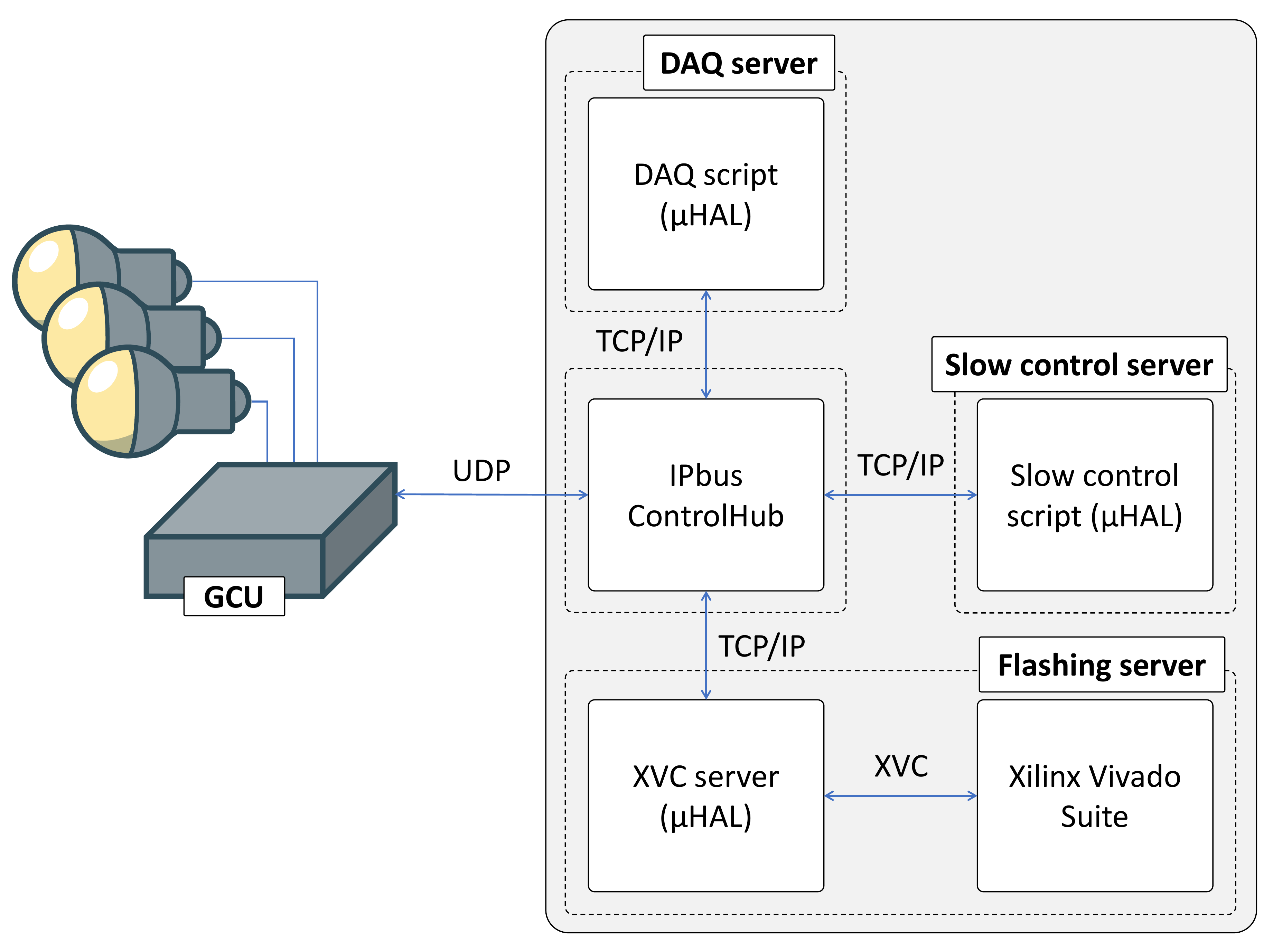}
  \caption{\label{fig:server-scheme}Scheme of the IPbus implementation within the DAQ infrastructure. The $\mu$HAL clients serving the DAQ, slow control and reprogramming capabilities communicate through the TCP/IP protocol with the ControlHub, which manages the requests to/from the GCUs via UDP. A custom Xilinx Virtual Cable (XVC) is used to route the Vivado Lab software connection to the GCU via Ethernet. The three $\mu$HAL components can operate within the same machine or in different machines, communicating via TCP/IP over Ethernet.}
\end{figure}

The server-side implementation of the IPbus is divided into three parts: (i) a high-performance C++ $\mu$HAL client for the acquisition of data, (ii) a Python $\mu$HAL client for the low-frequency monitoring of the GCU slow control parameters and (iii) a $\mu$HAL client for the upgrade of the GCUs' firmware and the eventual debug of firmware errors. 
All three of these data streams pass through the ControlHub server to reach the GCUs (Figure \ref{fig:server-scheme}). The communication between the $\mu$HAL clients and the ControlHub occurs through the TCP/IP protocol, meaning that components can reside on the same machine or in different machines if necessary. During the mass testing, components operated on the same server, but they are envisaged to be deployed to different machines in the JUNO experiment.

\paragraph{Firmware flashing}
\label{jtag}
The Vivado Lab software suite \cite{vivado} is used to reconfigure and update the firmware of the Xilinx Kintex-7 boards. This software is used to manage and program the FPGA boards through a JTAG connection, a common hardware interface that provides a way to communicate directly with the chips on a board. This connection is also used to monitor debugging parameters pre-established during firmware development (e.g. number of bitflips in the sync link, IPbus commands sento to the GCU).
Since the only connection between the server and the GCUs is an Ethernet link, we rely on the TCP/IP-based Xilinx Virtual Cable (XVC) protocol to access and debug the FPGA over Ethernet. A custom XVC daemon installed on the server receives the requests from the Vivado Lab software and routes them to the GCUs via IPbus through the ControlHub. 

The Kintex-7 reprogramming procedure consists of two steps. First a \textit{programmer} is mounted on the volatile memory of the FPGA. This is a firmware that pings on a specific IP address and which will be used to write the firmware on the permanent flash memory of the Kintex-7. Once installed the \textit{programmer}, the FPGAprog utility \cite{fpgaprog} is used to write the new firmware to the board through the specific \textit{programmer}'s IP address. While the flashing of the \textit{programmer} has a great impact on the server resources, the programming of the permanent flash memory is extremely lightweight.

At the end of the procedure, the FPGA reboots automatically and loads the updated firmware from the flash memory. For a single GCU this procedure takes a few seconds when happening through a physical JTAG link, but can take up to 2 minutes when relying on a virtual JTAG connection over Ethernet. Since the JUNO experiment will employ more than 7000 GCUs, it was necessary to develop a procedure for a parallel and efficient reprogramming of the cards.

A workers-based system was therefore developed in bash\footnote{Bourne Again SHell}. A number of \textit{workers} equal to the number of available cores are initialized (i.e. 24 in our server). A \textit{job} is assigned to each request for programming a GCU and is inserted in a FIFO queue. As soon as a \textit{worker} is available, it takes the first \textit{job} from the FIFO to execute it. When the \textit{worker} finishes the \textit{programmer}'s flashing \textit{job}, it launches in the background a FPGAprog process for writing the firmware to the GCU's permanent flash memory before moving on to the next \textit{job}. At the end of the procedure, the flashing script verifies that each GCU booted the correct firmware version. In case the GCU firmware update failed, a new \textit{job} is resubmitted to the FIFO queue. 

% \begin{figure}[hbtp] \centering
%   \includegraphics[width=0.9\columnwidth]{figures/scheme_flashing.pdf}
%   \caption{\label{fig:flash-scheme}Scheme of the flashing procedure.}
% \end{figure}

\begin{figure}[hbtp] \centering
  \includegraphics[width=1.\columnwidth]{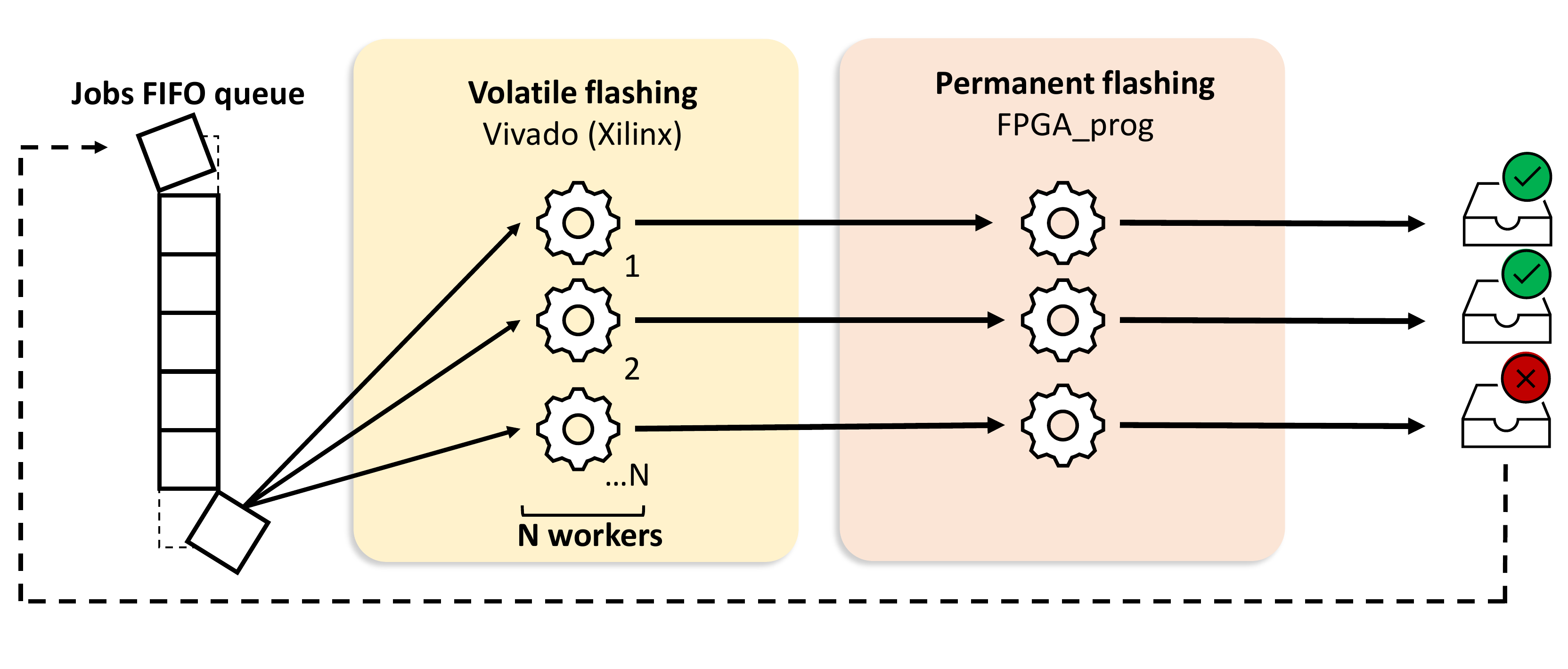}
  \caption{\label{fig:flash-scheme}illustrative scheme of the procedure adopted for the parallel reprogramming of GCUs.}
\end{figure}

\paragraph{Slow control}
% \textbf{[TO DO:] Script python per la lettura dei parametri di XXX. Ci sono dei client uHAL che fanno richieste routed attraverso il controlhub. Questi parametri vengono inseriti in un DataBase?} \\
The IPbus protocol is exploited to access and monitor a variety of internal parameters and sensors present on the GCU and keep track of the board's overall status.
The slow control monitoring runs in parallel to the DAQ, sharing the same IPbus transport layer thanks to the ControlHub capabilities. During the slow control, the following parameters are monitored for each GCU: temperature of the FPGA, temperature of each HVU, high voltage value of each HVU, FPGA internal supply voltage, supply voltage for FPGA RAM memorires, FPGA auxiliary supply voltage, external reference voltages for FPGA internal ADC. 
% All monitored parameters are stored in a dedicated database, queryable to investigate the stability of the GCUs over time. Any alarming value of slow control parameters (e.g., unusually high temperatures) causes a notification that can be used to take prompt actions (e.g., turn off the power supply) and prevent any damage to the instrumentation. Additional information about the slow control implementation can be found in \cite{bib:juno:kunshan_tests}.

\begin{figure}[hbtp] \centering
  \includegraphics[width=1\columnwidth]{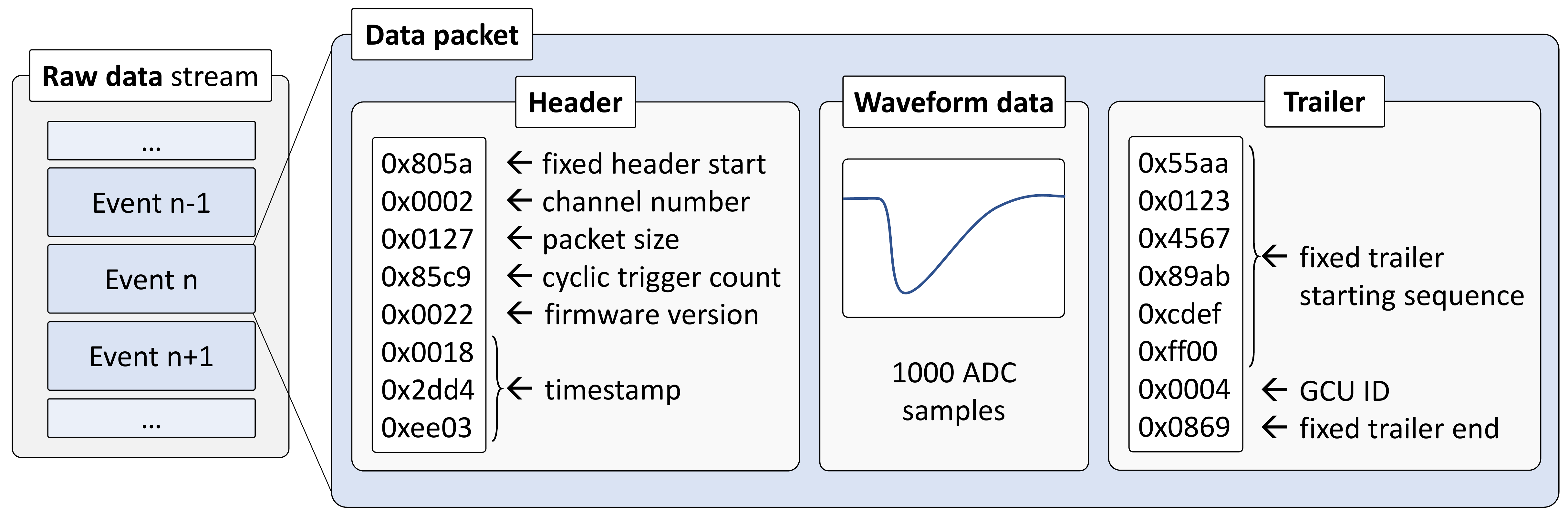}
  \caption{\label{fig:packet}general structure of the raw data. Every event is composed of header and trailer sequences wrapped around the waveform data. 
%   The \textit{timestamp} and the \textit{trigger window} are here highlighted.
}
\end{figure}

\paragraph{Data acquisition}
% \textbf{[TO DO:] Qui dobbiamo parlare di single, della blocksize e del fatto che ci sia un client per ogni GCU. Come funziona single? Cosa succede veramente alle richieste? Si legge dalla FIFO? Si legge da un buffer IPbus?} \\
For each GCU, a DAQ script based on a C++ $\mu$HAL client is initialized for the acquisition of data. The DAQ script queries via an IPbus transaction the IPbus DAQ on board of the GCU. For each request, the script specifies the size of the chunk of data to transfer (i.e. the \textit{DAQ Buffer Size}). Whenever the GCU has enough data to fill the requested buffer, it transfers the data to the server and empties the GCU's FIFO. The FIFO occupancy value parameter is exposed to IPbus, opening up to the possibility of complex adaptive scripts varying the requested block size as a function of the trigger rate and maximizing DAQ efficiency.
The \textit{DAQ Buffer Size} (BS) is the length parameter of the memory buffer in which the first chunk of data read by the DAQ is temporarily stored. The BS is expressed in units of $32 \ \si{bit}$s and its maximum value corresponds to the FIFO depth, that is $2^{13} \ \si{Byte}$s (i.e. 2048 32-bit blocks). The acquired data are stored in a binary file for each active channel (3 for each GCU) and are organized in data packets, each one having a variable number of \textit{words}, defined as sequences of 16 bits. Any packet is wrapped by an \textit{header} and a \textit{trailer} 8-words sequences, that provide unique information on each event (Figure \ref{fig:packet}).
%
% \bigskip
% %
% \begin{minipage}{.5\textwidth}
%     \centering
%     \begin{tabular}{|c|l}
% \cline{1-1}
% \textit{Header}  & \\
% \cline{1-1}
% \verb!0x805a! & $\leftarrow \text{Fixed start}$ \\
% \verb!0x0002! & $\leftarrow \text{Channel number (0,1,2)}$\\
% \verb!0x0040! & $\leftarrow \text{Trigger window}$\\
% \verb!0x85c9! & $\leftarrow \text{Trigger count}$ \\
% \verb!0x0022! & $\leftarrow \text{Firmware version}$ \\
% \verb!0x0018! &  \multirow{3}{*}{$\leftarrow \text{Timestamp}$}\\
% \verb!0x2dd4! & \\
% \verb!0xee03! & \\
% \cline{1-1}
%     \end{tabular}
% \end{minipage}%
% \begin{minipage}{0.5\textwidth}
%     \centering
%     \begin{tabular}{|c|l}
% \cline{1-1}
% \textit{Trailer}  & \\
% \cline{1-1}
% \verb!0x55aa! & \multirow{6}{*}{$\leftarrow \text{Fixed starting sequence}$}  \\
% \verb!0x0123! & \\
% \verb!0x4567! & \\
% \verb!0x89ab! & \\
% \verb!0xcdef! & \\
% \verb!0xff00! & \\
% \verb!0x0002! & $\leftarrow \text{GCU ID}$\\
% \verb!0x0869! & $\leftarrow \text{Fixed end}$\\
% \cline{1-1}
%     \end{tabular}
% %
% \end{minipage}
% %    
% \bigskip
%
%
Apart from the fixed sequences, each header is composed of the GCU channel number, the packet size, the trigger count and the firmware version. The packet size indicates the size of the transferred data packet including header and trailer in units of 8 words. 
% Therefore, in order to express the \textit{waveform size} in $\si{ns}$, one subtracts the TW by 2 and then multiplies it by 8. 
The trigger count is a hexadecimal cyclic number, that can be exploited for debugging. Finally, the last 48 bit sequence is the \textit{timestamp}, which gives the time reference in units of 8 ns. The trailer is composed of a fixed starting sequence, the GCU ID number and a fixed ending word. The actual event data is stored as a sequence of words given by the packet size minus header and trailer and stores the digitized waveform, which starts from the reference timestamp found in the header.

\section{IPbus performance assessment}
\label{sec:perfomance-tests}

% ---------------------------------------------------
% FIRMWARE FLASHING PERFORMANCE 
% ---------------------------------------------------
\subsection{Firmware flashing}
\label{subsec:firmware_flashing}
\begin{figure}[hbtp] \centering
  \includegraphics[width=1\columnwidth]{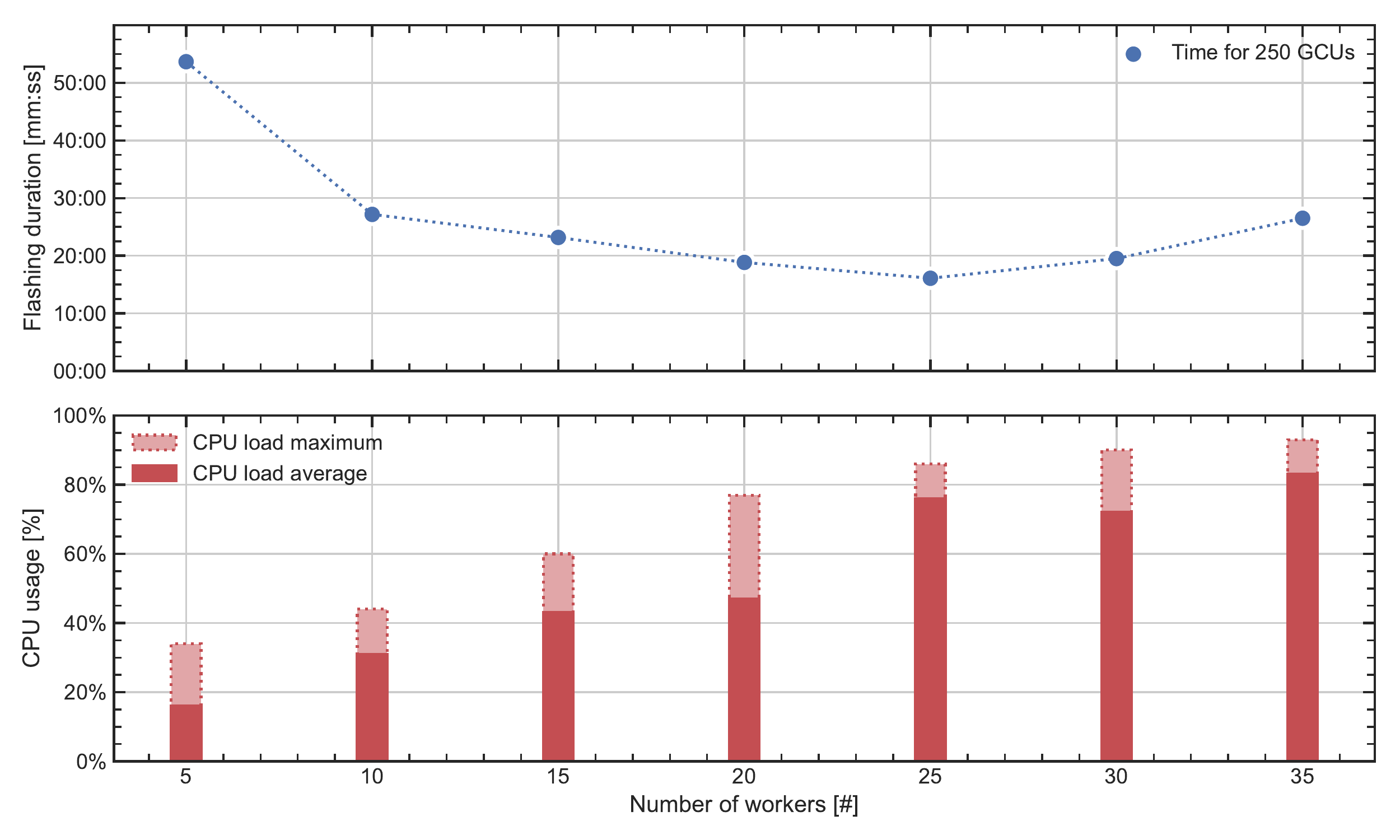}
  \caption{\label{fig:perf2}Duration (top) and corresponding CPU usage log (bottom) for the reprogramming procedure of 250 GCUs.}
\end{figure}

In order to assess the performances of the IPbus based implementation of the reprogramming framework, the time needed to flash a new firmware to a set of 250 GCUs in the Kunshan test setup \cite{bib:juno:kunshan_tests} was measured. The procedure was repeated with different configuration, initializing from time to time a different number of parallel workers capable of handling the flashing jobs. Simultaneously, the server resources allocated to the flashing procedure were monitored (Figure \ref{fig:perf2}).
% \ref{fig:perf1}).

% \begin{figure}[hbtp] \centering
%   \includegraphics[width=1\columnwidth]{figures/performances1.pdf}
%   \caption{\label{fig:perf1}XXX.}
% \end{figure}

The optimal configuration is obtained when initializing a number of workers approximately equal to the number of available cores on the machine (i.e. 24 cores). This configuration yields the minimum time needed to reprogram the complete set of GCUs, while exploiting most of the resources provided by the server. In this configuration it is possible to update the firmware of 250 GCUs in about 15 minutes. 
%Since the number of GCUs that each DAQ server will manage in the experiment is lower than 250 GCUs/server, this means that in JUNO it will be possible to completely reprogram the entire FE in about 15 minutes.

% ---------------------------------------------------
% SLOW CONTROL PERFORMANCE 
% ---------------------------------------------------
% \subsection{Slow control}
% \label{subsec:slow_control}
% \input{source/SlowControl}

% ---------------------------------------------------
% DAQ PERFORMANCE 
% ---------------------------------------------------
\subsection{Data acquisition}
\label{subsec:daq}
In order to assess the IPbus implementation performances in the data acquisition, several measurements with different trigger rates, different DAQ and network parameters and different number of GCUs run in parallel have been performed.
Two different metrics useful for the evaluation of the DAQ performances have been identified: (i) the maximal transferred bandwidth, that is the maximum amount of data transferred from the GCU to the server, and (ii) the survival fraction (or efficiency), that is the number of acquired events over the number of expected triggers. The tests were carried out using an external global trigger with fixed frequency provided by the CTU.

The analysis is carried out on the raw data saved to the server. Data are temporarily stored to ram-disk during acquisition and then moved to disk at the end of the test. For each acquired channel, a dedicated software is used to verify the validity of each event and to extract the timestamp and the metadata of that event.
When a channel misses a timestamp or the event data packet is corrupted, the corresponding event is considered lost. 
The number of acquired events is hence calculated by counting the number of timestamps obtained for each channel. The corresponding bandwidth can be calculated estimating the total amount of data transferred (i.e. the number of events multiplied by the size of each event, 2032 bytes\footnote{1 waveform packet = 16 header bytes + 16 trailer bytes + 1 $\mu$s x 1000 ADC/$\mu$s x 2 bytes/ADC = 2032 bytes}) divided by the acquisition time (i.e. the difference between the last and first timestamps acquired).
The number of expected events, on the other hand, is calculated by multiplying the trigger rate by the acquisition time. In this regard, since all tests have been performed by using an external trigger with fixed frequency, the trigger rate is extracted by the timestamp distribution with an arithmetic mean of the timestamp differences between consecutive events. The error attributed to the frequency is derived from the standard deviation of the timestamp differences.
The uncertainty on the expected number of events can be obtained by propagating the error on the trigger rate; the error on the number of registered events is estimated to be poissonian.

\begin{figure}[hbtp] \centering
  \includegraphics[width=1.
  \columnwidth]{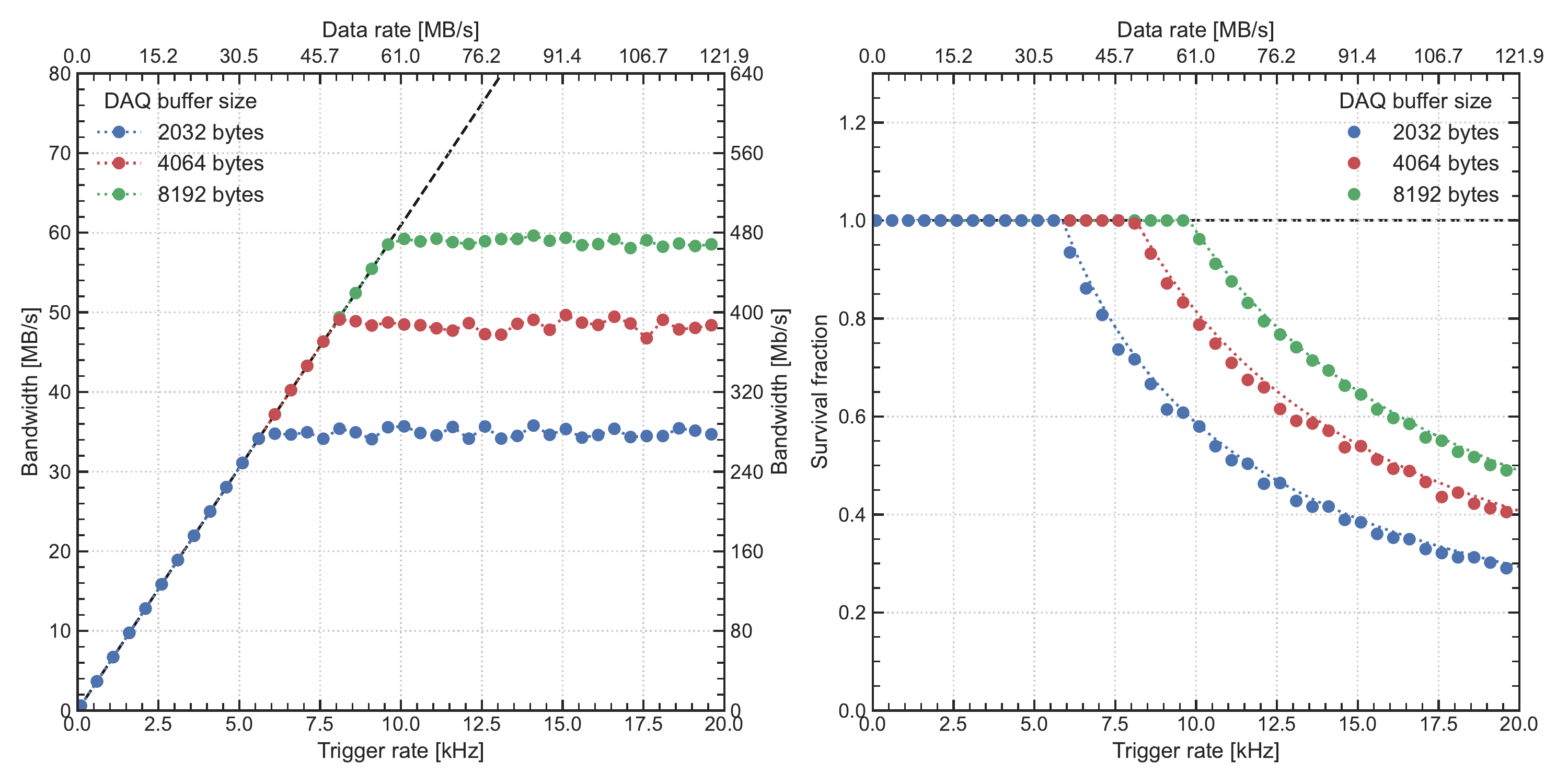}
  \caption{\label{fig:bw}transferred bandwidth (left) and corresponding survival fraction (right) for a single GCU as a function of the trigger rate. Acquisitions performed with different DAQ buffer sizes are reported in different colors.}
\end{figure}

The first parameter ruling the maximal performances of the readout is the DAQ buffer size. For a single GCU, the optimal readout performances can be achieved by setting a DAQ buffer size equal to the GCU's onboard IPbus FIFO size. Lower sizes require a higher number of transactions to empty the FIFO, the additional latency limits the reachable maximal bandwidth. Bigger sizes yield higher maximal bandwidths. Once that the maximal bandwidth has been reached, the data readout saturates and part of the acquired events cannot be transferred to the DAQ server (Figure \ref{fig:bw}). This permits to reach for a single GCUs readout performances of about 60 MB/s, corresponding to a trigger rate of nearly 10 kHz, well above the design requirements of 1 kHz.

The stability of the readout setup and of the synchronous link has been verified through a 350-hours long acquisition performed in the LNL setup \cite{bib:proc:jelmini_2021}. A rate test of cosmic muons has been performed on 39 channels in parallel, using the coincidence of three plastic scintillator bars as an external trigger connected to the BEC. The resulting cosmic muon rate (about 3~Hz) remained stable over almost 14 days on all the 13 employed GCUs for the entire duration of the test, without any clock re-synchronization \cite{bib:FPGA}.

\begin{figure}[hbtp] \centering
  \includegraphics[width=1.\columnwidth]{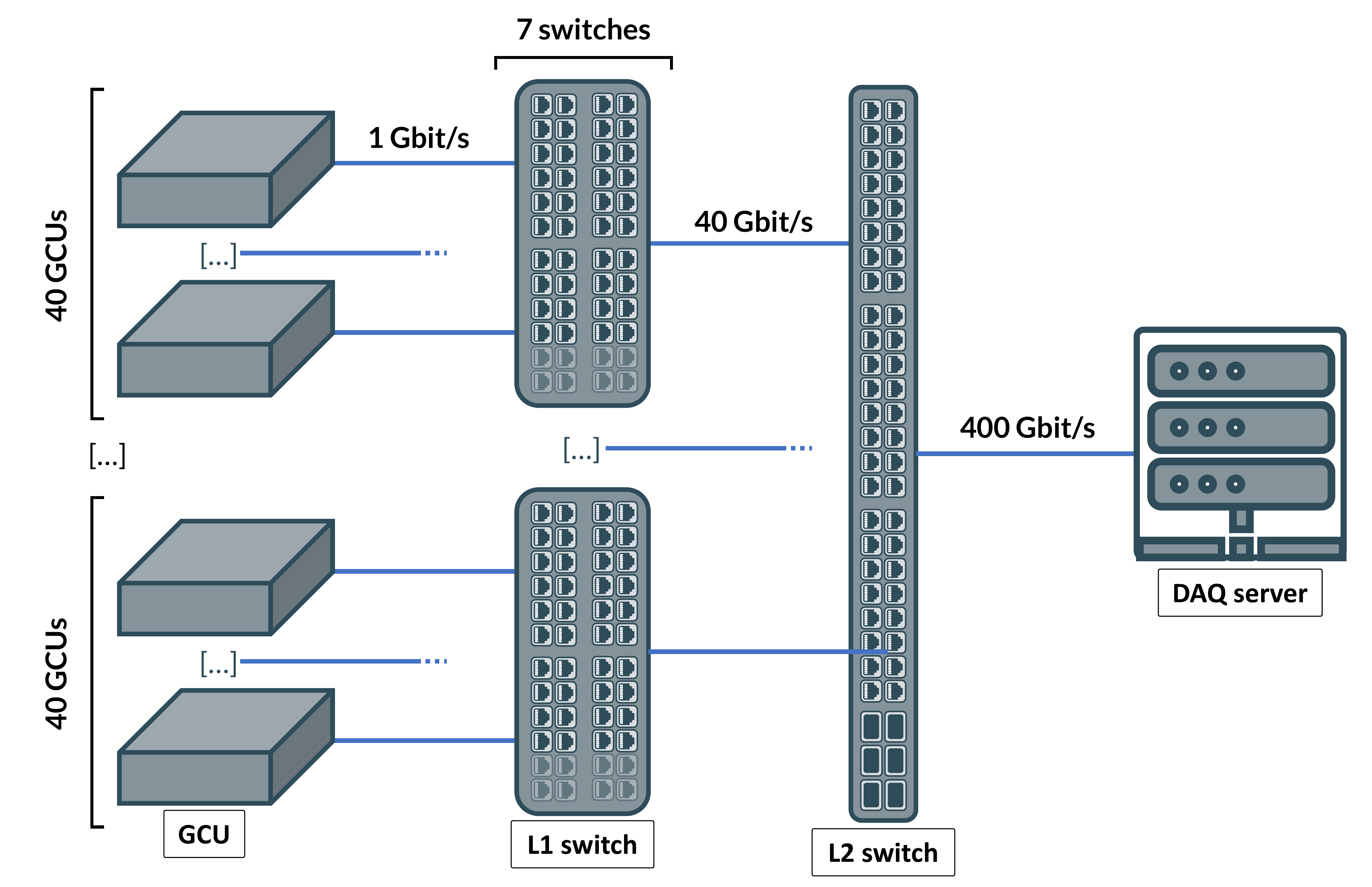}
  \caption{\label{fig:network}scheme of the network infrastructure in the Kunshan setup. Groups of 40 GCUs are connected to a L1 switch. A total of 7 L1 switches are then connected to a single L2 switch communicating with the DAQ server through a Terabyte Ethernet port.}
\end{figure}

\begin{figure}[hbtp] \centering
  \includegraphics[width=1.\columnwidth]{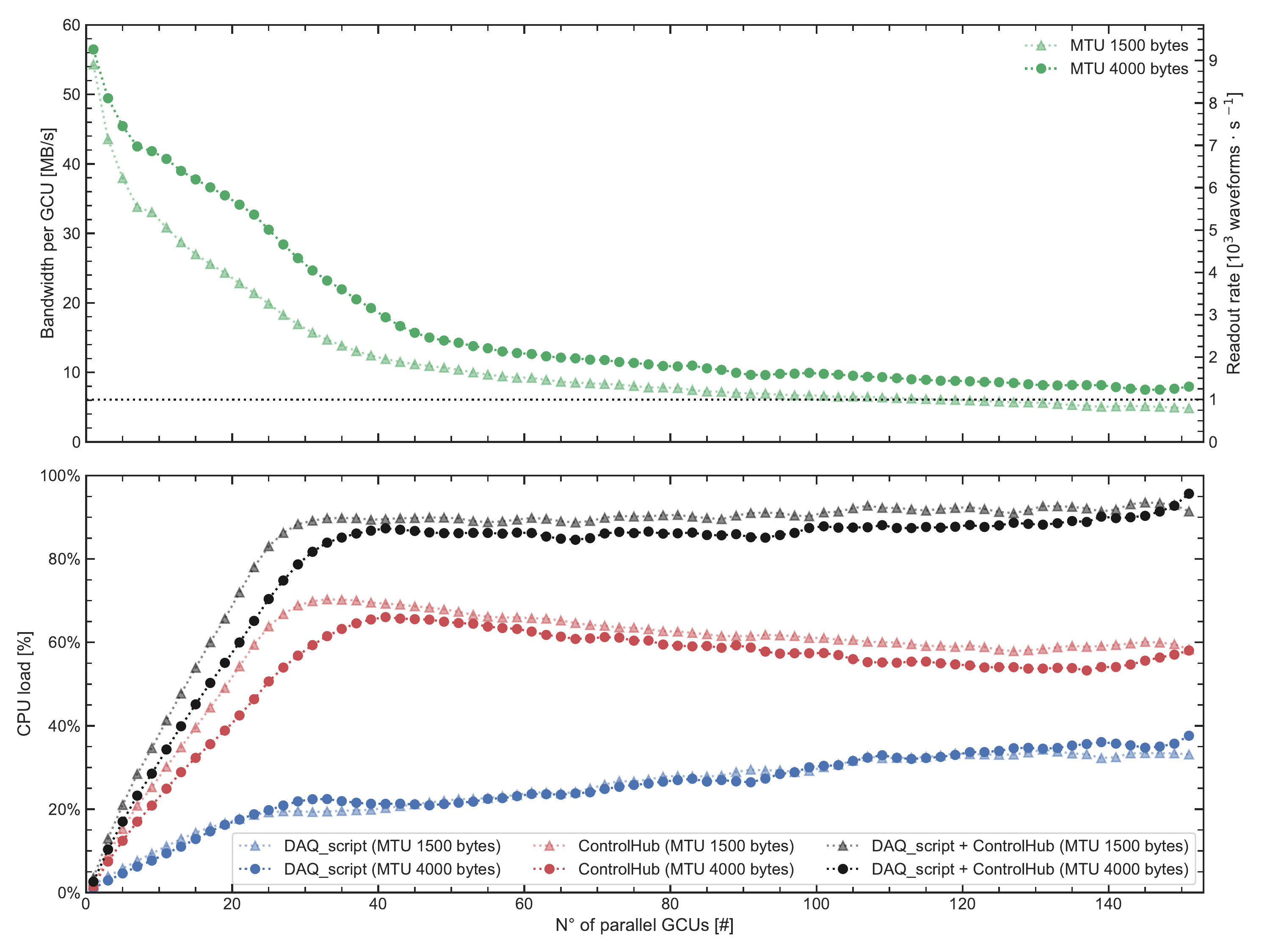}
  \caption{\label{fig:several-gcus}relative bandwidth trend (top) and corresponding CPU usage log (bottom) as function of the number of GCUs run in parallel. Each data point corresponds to a 60~s acquisition. The total CPU usage (in black) is broken down into its ControlHub (in red) and DAQ acquisition script (in blue) components. Acquisitions have been performed with different Maximum Trasmission Units (MTUs) set on the network interfaces, here represented by different levels of transparency.}
\end{figure}

\begin{figure}[hbtp] \centering
  \includegraphics[width=1.
  \columnwidth]{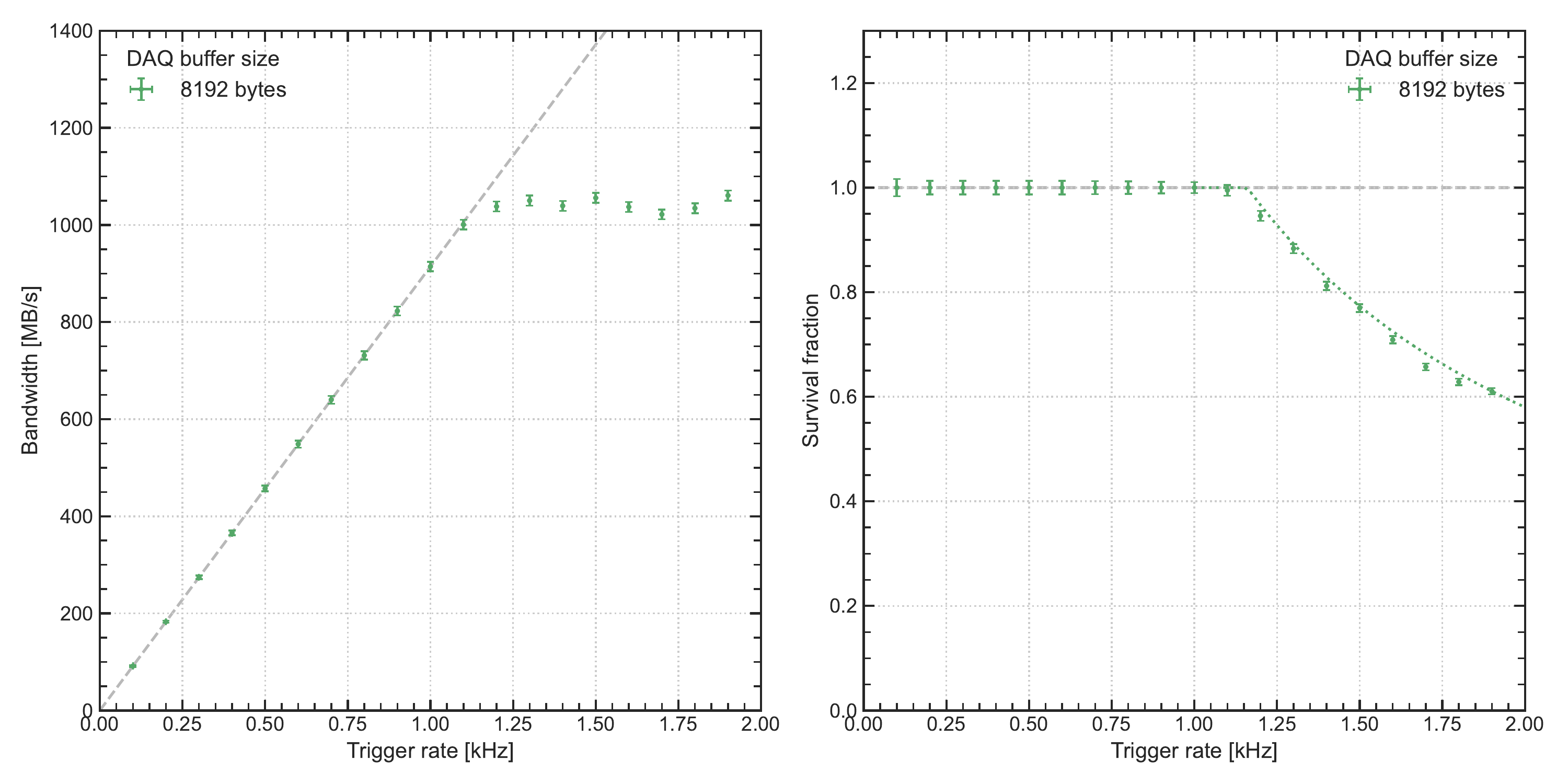}
  \caption{\label{fig:rate-test}transferred bandwidth and corresponding survival fraction for 150~GCUs run in parallel as a function of the trigger rate.}
\end{figure}

When dealing with a higher number of GCUs two possible bottlenecks may be due to (i) network performances and (ii) server resources. For what concerns the network, each GCU accommodates a 1 Gbit/s uplink connected to a 40 Gbit/s switch. In the Kunshan setup, a Terabyte Ethernet port assures transfer bandwidths up to 400~Gbit/s between GCUs and the server itself (Figure \ref{fig:network}). This is enough to accommodate for more than 800~GCUs per server reading out at peak bandwidths.
Server resources represent instead a strong constraint and need a thorough investigation. Several acquisitions with an increasing number of GCUs (from 1 to 150) read out in parallel have been performed (Figure \ref{fig:several-gcus}). Simultaneously, the server resources used by the DAQ processes during the acquisitions have been monitored.
The CPU usage rapidly increases up to 30 GCUs and then it settles up due to the CPU being saturated. The bandwidth transferred per GCU quickly decreases suggesting to be limited by the server resources. More specifically, most of the resources are occupied by the \texttt{ControlHub} server ensuring UDP reliability via IPbus. In order to meet the $1 \ \si{kHz}$ trigger rate requirement, the maximum number of GCUs read out per server has to be approximately 100. This result can be further improved, by means of the \textit{jumbo frame} feature implemented within the IPbus framework. Jumbo frames are Ethernet frames with up to 9000 Bytes of payload \cite{bib:ethernet-jumboframe}, overrunning the IEEE 802.3 standard of frames with payloads between 46 and 1500 Bytes. Generally, for one client controlling one device, the read or write latency for a single word is approximately $250 \ \si{\micro s}$ and the protocol becomes optimal by concatenating multiple transactions into each packet \cite{bib:ipbus-performance}.
The network interface implemented on the GCU permits to reach Maximum Transmission Units (MTUs) up to 4000 Bytes, against the standard 1500 Bytes. Increasing the MTU leads to a reduced number of network packets exchanged leading to a lower CPU usage by the \texttt{ControlHub}. The resources occupied by the acquisition itself do not change. Consequently, the relative bandwidth trend is shifted and this allows to reach higher rates without losing events or, equivalently, to employ up to 150 GCUs and still withstand the $1 \si{kHz}$ target trigger rate (Figure \ref{fig:rate-test}).

\section{Conclusion}
The JUNO experiment is expected to inaugurate a new precision era in the field of neutrinos. To reach its aims and the desired energy resolution, the experiment will make use of a total of 20,012 20-inch Large-PMTs. A key requirement for the success of JUNO is therefore the development of a read-out electronics capable of reliably managing, digitizing and acquiring the waveforms coming from the PMTs. This is achieved thanks to the employment of a custom dual-FPGA GCU system installed underwater in the proximity of the PMTs. 
Following CMS, ATLAS and ALICE successful implementation within their trigger systems, the IPbus protocol has been chosen as the designated transport layer for the readout of JUNO DAQ data and slow control parameters.
Two test facilities have been set up in Kunshan and Legnaro to test the performances of the IPbus-based readout of the Large-PMT electronics during the mass production. 
Carried out measurements showed that a single GCU is able to manage trigger rates up to about 10~kHz without any event loss, thus meeting and exceeding the JUNO design requirement of 1~kHz. When dealing with multiple GCUs run in parallel, the data acquisition performances are limited by the CPU resources of the DAQ server. After careful optimisation in the data acquisition and network parameters, it has been possible to manage with a single 24-cores server up to 150~GCUs, still meeting the required 1~kHz trigger rate. 
A dedicated parallelized software has been developed for an efficient and reprogramming of the GCUs's firmware via Ethernet. Performance tests performed on a 250 set of GCUs in Kunshan show that the developed procedure will permit to completely reprogram the JUNO front-end electronics in about 15 minutes.   

\section*{Acknowledgments}
Part of this work has been supported by the Italian-Chinese
collaborative research program jointly funded by the Italian Ministry of
Foreign Affairs and International Cooperation (MAECI) and the National
Natural Science Foundation of China (NSFC).

% \bibliography{references/juno_ipbus}
\bibliography{main}
%\bibliography{references/juno-ipbus} % da togliere
%\bibliography{references/juno_electests} % da aggiungere
%\bibliography{references/juno_ele_section} % da aggiungere

\end{document}